\documentclass[fleqn,usenatbib]{mnras}

\usepackage{newtxtext,newtxmath}

\usepackage[T1]{fontenc}

\DeclareRobustCommand{\VAN}[3]{#2}
\let\VANthebibliography\thebibliography
\def\thebibliography{\DeclareRobustCommand{\VAN}[3]{##3}\VANthebibliography}

\usepackage{graphicx}	
\usepackage{amsmath}	

\title[Primordial black holes in the Milky Way halo]
 {A critical analysis of the recent OGLE limits on stellar mass primordial
 black holes in the halo of the Milky Way}

\author[M. R. S. Hawkins and J. Garc\'ia-Bellido]{
M. R. S. Hawkins $^{1}$\thanks{E-mail: mrsh@roe.ac.uk}
J. Garc\'ia-Bellido $^{2}$\thanks{E-mail: juan.garciabellido@uam.es}
\\
$^{1}$Institute for Astronomy (IfA), University of Edinburgh,
 Royal Observatory, Blackford Hill, Edinburgh EH9 3HJ, UK\\
$^{2}$Instituto de F\'isica Te\'orica UAM/CSIC,
 Universidad Aut\'onoma de Madrid, 
 Nicol\'as Cabrera 13, Madrid 28049, Spain\\}

\date{Accepted XXX. Received YYY; in original form ZZZ}

\pubyear{2025}

\begin{document}
\label{firstpage}
\pagerange{\pageref{firstpage}--\pageref{lastpage}}
\maketitle

\begin{abstract}
This paper is a response to recent claims that a population of primordial
black holes in the Galactic halo has been ruled out by the OGLE
collaboration.  This claim was based on the latest results from the OGLE
microlensing survey towards the Large Magellanic Cloud which failed to
detect even the number of events expected from known stellar populations.
In particular, their results are completely inconsistent with the results
of the MACHO survey which detected a population of compact bodies in the 
Galactic halo which could not be accounted for by any known stellar
population.  The discrepancy between the results of these two groups has a
long history, and includes problems such as different choice of
photometric passbands, quality of light curves, microlensing event
selection, detection efficiency, self lensing and halo models.  In this
paper it is demonstrated that these issues not only account for the
discrepancy between the OGLE and MACHO results, but imply that the OGLE
observations can put no meaningful constraints on a population of
primordial black holes in the Galactic halo.
\end{abstract}

\begin{keywords}
quasars: general -- gravitational lensing: micro -- dark matter
\end{keywords}



\section{Introduction}
\label{int}

The idea that a large component of the Universe is in some form dark
matter goes back to the early 20th century.  It was soon realised that
this dark material must be non-baryonic, and a number of possible
candidates were considered in the literature.  The widespread belief
developed that dark matter was in the form of some as yet undiscovered
subatomic particle, but a number of more massive candidates such as cosmic
strings or primordial black holes (PBHs) were also considered.  An
ambitious idea to test these alternatives was proposed by \cite{p86}, that
involved the photometric monitoring of several million stars in the Large
Magellanic Cloud (LMC) to look for rare microlensing events in the light
curves which might be associated with a population of compact bodies in
the halo of the Milky Way acting as microlenses.  The statistics of the
detections would then be used to determine whether the compact bodies
could make up the dark matter.

The results of this project \citep{a00} are well known, and constituted an
outstanding success.  The MACHO collaboration reported 13-17 microlensing
events, depending on the selection criteria, of which they estimated that
no more than 3 or 4 were attributable to the known stellar background in
the form of lenses in the Galactic disc or self-lensing events in the LMC.
In spite of this technical success, and of apparently detecting a new and
unidentified population of compact objects populating the Galactic halo,
the MACHO collaboration were at pains to point out that they did not
consider that the detections implied enough mass to make up the halo dark
matter.  This was based on their adopted model of the Galaxy \citep{a96}
which featured a flat rotation curve and a massive halo.  At this time,
little was known about the rotation curve of the Milky Way \citep{g24},
and the adoption of a flat rotation curve seems to have been based on
analogy with nearby bright spiral galaxies where the rotation curves could
be quite easily measured. This misunderstanding persisted for 15 years
until it became possible to make direct measurements of the Galactic
rotation curve \citep{x08,s13,b14}, which were sufficient to show that the
circular velocity of the Galalaxy actually decreases out to the distance
of the LMC \citep{h15}.  This finding continues to be confirmed as better
data become available \citep{c18,g24}.

The important point about the observation of the declining rotation curve
is that it implies a reduced halo mass density, and hence a smaller
surface density in dark matter towards the LMC.  For a halo made up of
PBHs this implies a smaller optical depth to microlensing $\tau$ which is
now consistent with the value of $\tau$ observed by the MACHO
collaboration \citep{a00} for models of the Galaxy constrained by the
observed rotation curve \citep{h15}.

Given the importance of these observations, which essentially identify
primordial black holes as at least the major component of dark matter, it
is fortuneate that two other groups have taken up the challenge to
reproduce the MACHO results.  We have recently discussed the EROS project
in some detail as part of a paper incorporating the GAIA observations into
the measurement of the Galactic rotation curve \citep{g24}.  Our present
paper has been prompted by the widespread attention given to the recent
paper in {\it Nature} by \cite{m24a}, with further more specific arguments
in an accompanying paper \citep{m24b}.  Taken together, these papers are
presented as ruling out primordial holes as dark matter, a claim which we
strongly contest.  We are concerned that these papers have been widely
taken at face value, and it is our intention here to present the reasons
why the constraints are not valid, and to make the case that primordial
black holes have indeed been detected in the Milky Way halo.

\section{The OGLE project}
\label{ogl}

The OGLE project is in overall concept very similar to the MACHO project,
involving the photometric monitoring of several million stars in the
Magellanic Clouds to look for microlensing events due to compact bodies in
the Milky Way halo.  There are however important differences in the
construction of the two surveys which have the potential to result in very
different detection rates for microlensing events.  These differences
include the total number of source stars monitored, the surface density
distribution of the LMC source stars, the photometric passbands adopted
for the monitoring programme, the limiting magnitude for each passband,
and the estimation of the detection efficiency for microlensing events.
In addition, differences in the choice of Galaxy models also affect the
expected number of microlensing events and the likelihood of self lensing.

The first phase of the OGLE project (OGLE-II) \citep{w09} involved
monitoring about 5 million stars in the LMC from the years 1996 to 2000,
covering roughly the same period as for the 12 million stars observed by
the MACHO survey. There were however important differences between the
two surveys. The OGLE observations were made through standard $V$ and
$I$ band filters in contrast to the MACHO survey which used a dichroic
beam splitter and filters to divide the light into "blue" and "red"
passbands.  These two passbands are both substantially bluer than the $V$
and $I$ bands used by OGLE, which has major consequences for the detection
of LMC microlensing events which we discuss further below.  The OGLE-II
survey covered a total of around 5 deg$^2$, mainly including the densest
star fields in the bar of the LMC.  The fields were monitored in the
$I$-band on average every third night, with observations in the $V$-band
every 11th night.  The detection of microlensing events was restricted to
the $I$-band observations, meaning there was no colour information to aid
in the classification of photometric features in the light curves.  This
contrasts with the 15 deg$^2$ of the MACHO survey, which included a number
of relatively sparsely populated star fields, and was monitored
simultaneouly in the red and blue passbands.  The end product of this
first phase of the OGLE microlensing survey was the identification of just
two light curves satisfying the OGLE-II selection criteria \citep{w09} for
a candidate microlensing event.  This is to be compared with the 13-17
events detected by the MACHO survey.

These results in themselves suggested a discrepancy between the two
surveys, but two further steps were necessary to estimate the MACHO
content of the Galactic halo.  The first step was to model the detection
efficiency to give the ratio of detected events to those events actually
occuring within the observational limits of the survey.  In sparse stellar
fields this is a fairly straightforward calculation, allowing for down
time of the telescope, events outside the photometric limits of the
survey, incomplete light curves and so on.  However, in crowded star
fields the estimation of efficiency becomes extremely hard.  Overlapping
star images will distort the shape of the Paczy\'{n}ski curve which is the
primary criterion for identifying a microlensing event.  Similarly, source
stars merged with neighbouring stars of different colours will not show
achromatic variation in their light curves which is the second most
important requirement for detecting microlensing events.  Thirdly, changes
in seeing from night to night will mean that source stars can remain
single or merge at random with neighbouring stars on a nightly basis,
resulting in large unpredictable changes to the Paczy\'{n}ski profile.
An interesting insight into this issue can be gained from examination of
Figure 4 of \cite{a01}, where a Hubble Space Telescope image of a the
field around a MACHO microlensing event shows 5 star images within the
point spread function of the MACHO telescope.  When faced with this
problem, the MACHO project opted for a complex variation on a Monte Carlo
process, in which artificial events were injected into observed light
curves, and the resulting detection efficiency estimated.  This procedure
was first described by \cite{a96}, with subsequent major upgrades in
following papers \citep{a97,a00}.  The description of the estimation of
detection efficiency by \cite{w09} is brief, but appears to follow the
approach taken by the MACHO group.  Differences in detection efficiency
are unlikely to resolve the discrepancy between the MACHO and OGLE
results, but a reliable estimation is essential for accurately measuring
the value of $\tau$.  On the basis of their two microlensing event
detections, this is given by OGLE-II \citep{w09} to be
$\tau = 0.43 \pm 0.33 \times 10^{-7}$, about a third of the MACHO value of
$\tau = 1.2 \pm 0.4 \times 10^{-7}$ \citep{a00}.  Using the MACHO groups
favoured Model S for the Galactic halo, this translates into a MACHO halo
fraction of 20\% for the MACHO group and 8\% for OGLE-II.  It should
however be emphasised that the halo model upon which these figures are
based has now been ruled out by more recent observations \citep{g24}.
However, the final conclusion of the OGLE group was that their detections
were most likely to be self lensing by LMC stars \citep{w09}, quoting
estimates from the literature of around $\tau = 0.4 \times 10^{-7}$ for
the optical depth for self lensing in the LMC.

The second phase of the OGLE project (OGLE-III) \citep{w11} extended the
monitoring period for a further 8 years from 2001 until 2009, and
increased the sky coverage from 5 deg$^2$ to 40 deg$^2$, including
sparsely populated fields surrounding the LMC.  This had the effect of
increasing the number of stars monitored from 5 million to 35 million,
with the expectation that the number of events detected would rise to
around 20.  The main change between the two phases of the survey was the
installation of a new wide field camera, but the basic procedures of the
survey remained largely unchanged.  In particular, the event detection was
still based solely on $I$-band observations.

The main result of OGLE-III was the detection of two microlensing events
with the automated pipeline, and two other events found by visual
inspection.  The latter two were subsequently rejected as being
implausible microlensing events, leaving two acceptable microlensing
candidates.  This implied an optical depth to microlensing
$\tau = 0.15 \times 10^{-7}$, about a third of that expected for self
lensing, and would appear to render the identification of the two events
as self lensing somewhat insecure.  Indeed, the inability to detect
predicted self lensing events must raise doubts about the viability of
the whole OGLE event detection procedure, and especially the estimates of
the detecion efficiency.

The final phase of the OGLE project (OGLE-IV) \citep{m24b} extended
observations from 2010 to 2020, and increased the survey area to
300 deg$^2$ and the number of stars monitored to 62 million.  In order
to increase the length of the light curves, the observations for OGLE-III
were extended with OGLE-IV observations to give a baseline of 20 years.
Apart from installing a new mosaic CCD camera, and the photometric
corrections required to reduce the OGLE-III and OGLE-IV observations to a
combined system, the new survey was based on the same methodology as
earlier OGLE phases.  This included the continued use of $I$-band only
observations for the detection of microlensing events, and a similar
approach to estimating the detection efficiency.  The main result of this
extended survey was the detection of 13 candidate microlensing events
implying an optical depth to microlensing
$\tau = 0.121 \pm 0.037 \times 10^{-7}$.  Interestingly, this is still
only about a third of the expected self lensing events and again raises
questions about the effectiveness of the OGLE event detection procedures.

\section{Analysis of the OGLE results}
\label{res}

\begin{figure}
\centering
\begin{picture} (150,440) (45,-10)
\includegraphics[width=0.49\textwidth]{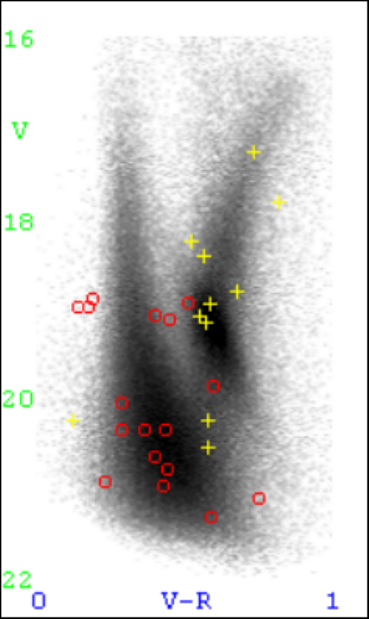}
\end{picture}
\caption{Hess diagram for LMC stars adapted from Alcock et al. (1999).
 The red circles show the positions of the microlensing candidates from
 Figure 7 of Alcock et al. (2000).  The yellow crosses show the positions
 of the microlensing candidates from Figure 5 of Mr\'{o}z et al. (2024),
 but with the $I$ and $(V-I)$ colours transformed to $V$ and $(V-R)$ to
 enable comparison of the two datasets.}
\label{fig1}
\end{figure}

The publication of the MACHO results by \cite{a00} reporting the detection
of 13-17 microlensing events in the Galactic halo implied a hitherto
unknown population of compact stellar mass bodies which could not be
accounted for by any known population of stars.  The imminent publication
of the results of a similar survey by the OGLE group was thus seen as a
necessary step to confirming such an important discovery.  However, it is
clear from Section~\ref{ogl} above that the first phase of the OGLE
observations in the LMC did not support the MACHO findings.  The detection
of only two events which they claimed was consistent with self lensing was
broadly speaking confirmed by their subsequent work, which is perhaps not
surprising as their observing and analysis strategy did not change
significantly, but only the area of sky and number of stars observed.
It is the purpose of this Section to highlight the main diferences between
the OGLE and MACHO surveys, with a view to understanding why their results
were so different.

\subsection{Photometric passbands}

It is correctly stated by \cite{w09} that observations of their LMC fields
were made in the $I$ and $V$ bands, but what is not so obvious without
careful reading of the paper is that only a quarter as many $V$ as $I$
frames were taken, and that the detection of microlensing events was
performed using $I$-band data only.  This lack of colour information has a
number of ramifications, and goes a long way to explaining the difference
between the MACHO and OGLE results.  To understand this it is necessary to
compare the distribution in colour and magnitude of the MACHO
microlensing candidates illustrated in Figure 7 of \cite{a00} with the
OGLE microlensing candidates illustrated in Figure 5 of \cite{m24b}.  To
facilitate this comparison, Fig.~\ref{fig1} shows a Hess diagram adapted
from  Figure 17 of \cite{a99} for LMC stars in MACHO field 11.  The MACHO
and OGLE candidates are superimposed on it as red circles and yellow
crosses respectively.  It will be seen that the source stars for the OGLE
detections are almost entirely confined to the red giant branch, whereas
the MACHO source stars are almost entirely confined to the top of the main
sequence.  To understand this remarkable difference, a number of factors
come into play affecting the likelihood of microlensing events being
detected.

We first note that no useful magnitude completeness limit is given for
either survey, which is not unreasonable given the problem of overlapping
images in crowded fields.  However, examination of the colour magnitude
diagrams suggests that the effective limit for the OGLE observations is
$I \sim 20$.  For blue stars this implies a limit of $V \sim 20$, which
would eliminate most main sequence stars in the LMC.  The giant and
supergiant branches however, with $V-I \sim 1$, will not be constrained by
the $I$-band limit.  This means that the OGLE sample of microlensing
sources is likely to be largely made up of red giants and supergiants,
which is what is observed.  However, for the MACHO sources, the effective
limit appears to be $V \sim 21$ allowing a large sample of stars at the
top of the main sequence as potential sources.  These stars will be
compact with radii much smaller than the Einstien radius of a stellar mass
lens, and should effectively act as point sources.

\begin{figure*}
\centering
\begin{picture} (0,120) (250,-10)
\includegraphics[width=0.32\textwidth]{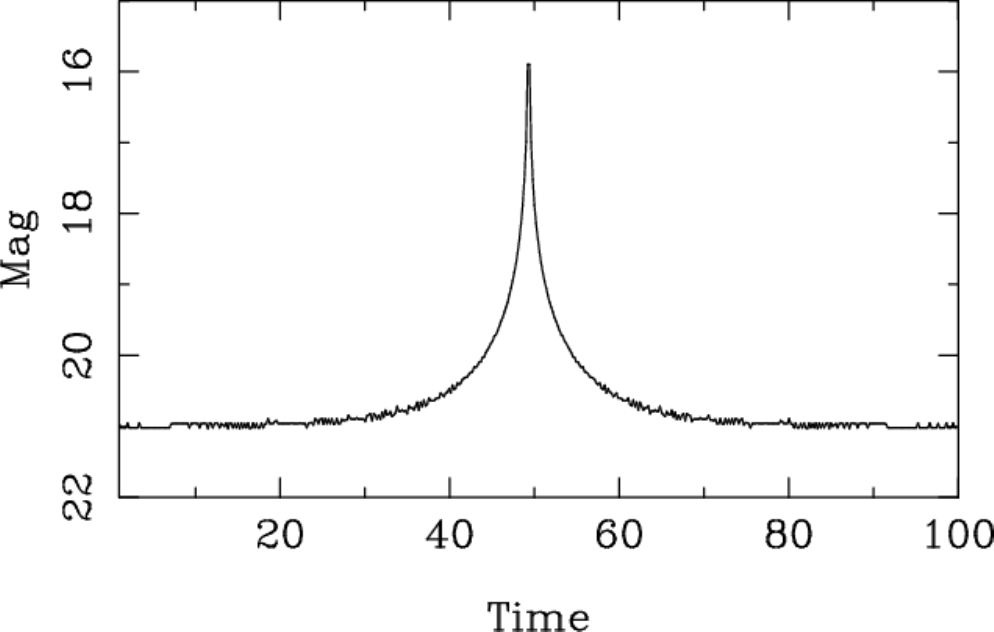}
\end{picture}
\begin{picture} (0,0) (80,-10)
\includegraphics[width=0.32\textwidth]{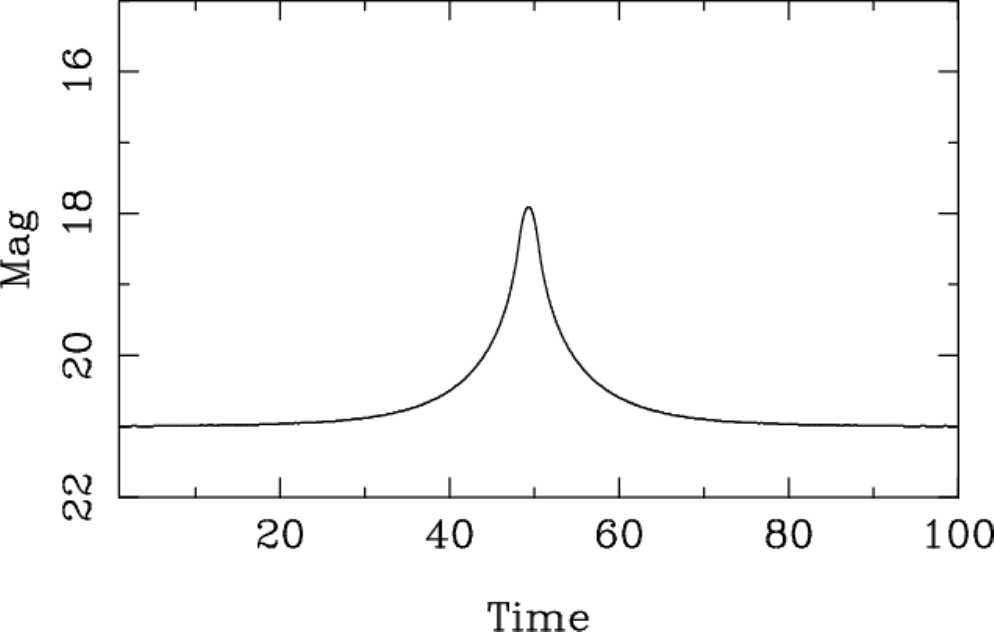}
\end{picture}
\begin{picture} (0,0) (-90,-10)
\includegraphics[width=0.32\textwidth]{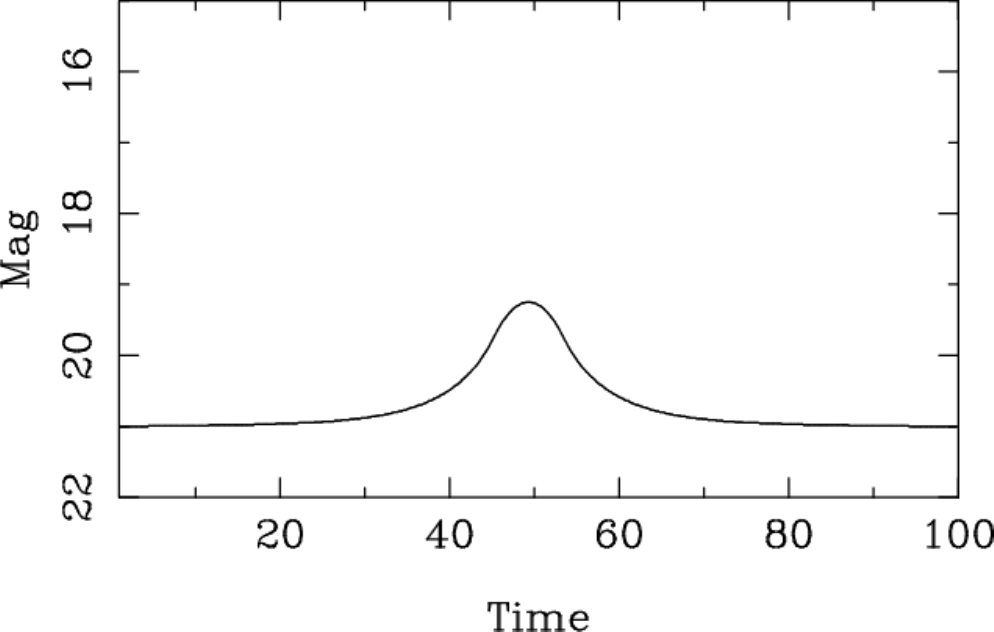}
\end{picture}
\caption{Simulated light curves for a point source (left), and source
 radius a tenth (centre) and a third (right) of the Einstein radius of
 the lens.}
\label{fig2}
\end{figure*}

The situation is very different for the giants and supergiants which are
the only sources observable in large numbers by OGLE.  The Einstein radius
of a stellar mass lens in the LMC is around $10^{14}$ cm with only weak
dependence on lens mass.  The radius of the faintest red giants is around
$10^{12}$ cm which is small enough for the star to approximate a point
source, and these stars appear to make up the bulk of the OGLE new
detections.  However, a substantial part of the LMC colour magnitude
diagram is populated by much more luminous giants and supergiants, with
radii ranging up to $3 \times 10^{13}$ cm, or around a third of the
Einstein radius of the lens.  The effect of increasing the source size or
stellar disc has been treated analytically by \cite{w94}, and is
illustrated in Fig.~\ref{fig2} from computer simulations, which shows the
difference betweeen the Paczy\'{n}ski profile from a standard point source
model, and the observed profile when convolved with a finite source size.
The effects of limb-darkening on microlensing light curves have also been
investigated by \cite{w19}, and their Figure 2 illustrates the resulting
large departures in microlensing amplification from a uniformly
illuminated disc.  This will result in the observed profile not being
recognized as a Paczy\'{n}ski profile, and the signal-to-noise of a
microlensing event being very much reduced.  In addition, the absence of
colour information in the light curve will exacerbate the difficulty of
recognising any feature as a microlensing event.

\subsection{Quality of light curves}

The production of millions of light curves in the LMC for the detection of
microlensing events was a Herculean task for both the MACHO and OGLE
collaborations.  However, due to differing observing strategies the
quality of the resulting observations, and their suitability for event
detection, was not the same.  We have noted above that observations in a
single passband with no colour information severely limit the likelihood
of microlensing detections.  For the MACHO survey, achromatic variation is
an important requirement for identifying a microlensing event, which
unfortunately was not available for the OGLE survey.  This is particularly
important in a crowded field where features in a light curve are likely to
be distorted by neighbouring stars.  The MACHO group also use colour
information to help identify supernova light curves, a source of
contaminent which does not appear to be mentioned by \cite{m24a}.  This
means that the classification of "bumps" as the OGLE group describe them
must rely almost entirely on the extent to which they resemble a
Paczy\'{n}ski profile.  A related issue to the problem of identifying
supernovae concerns the variability of source stars, especially red
giants.  Part of the OGLE event selection procedure is to reject any event
where the source shows signs of intrinsic variabilty, irrespective of
amplitude, magnification or cadence.  It can be seen from Fig.~\ref{fig1}
that most of the OGLE sample of microlensing candidates lie in the red
giant part of the colour magnitude diagram, and are likely to be variable.
This will result in a high probability that microlensing events in their
light curves will be ignored, even when intrinsic variability is only
detected after the completion of the survey.

\begin{figure}
\centering
\begin{picture} (150,200) (45,0)
\includegraphics[width=0.48\textwidth]{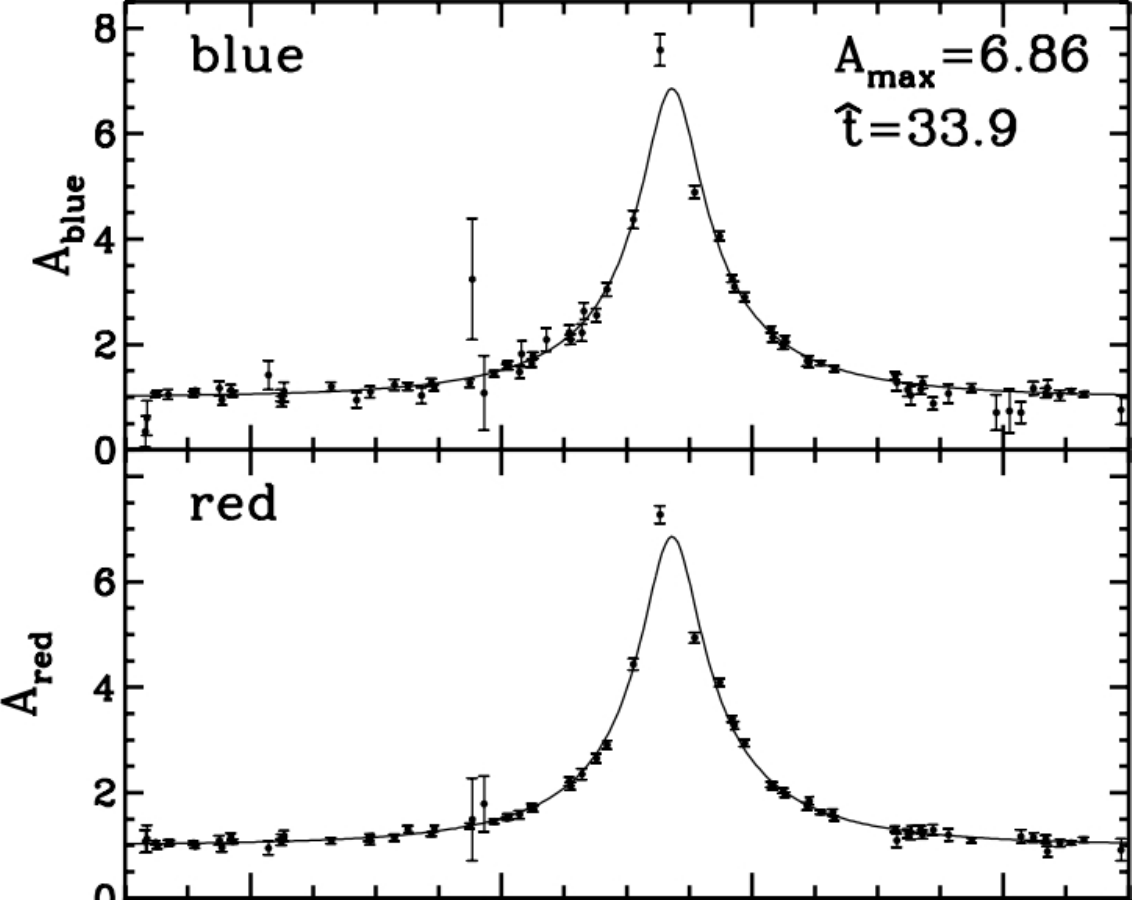}
\end{picture}
\caption{Light curves in the MACHO blue and red passbands adapted from
Figure 2 of Alcock et al. (1993) for a candidate microlensing event.}
\label{fig3}
\end{figure}

\begin{figure}
\centering
\begin{picture} (150,240) (45,-130)
\includegraphics[width=0.49\textwidth]{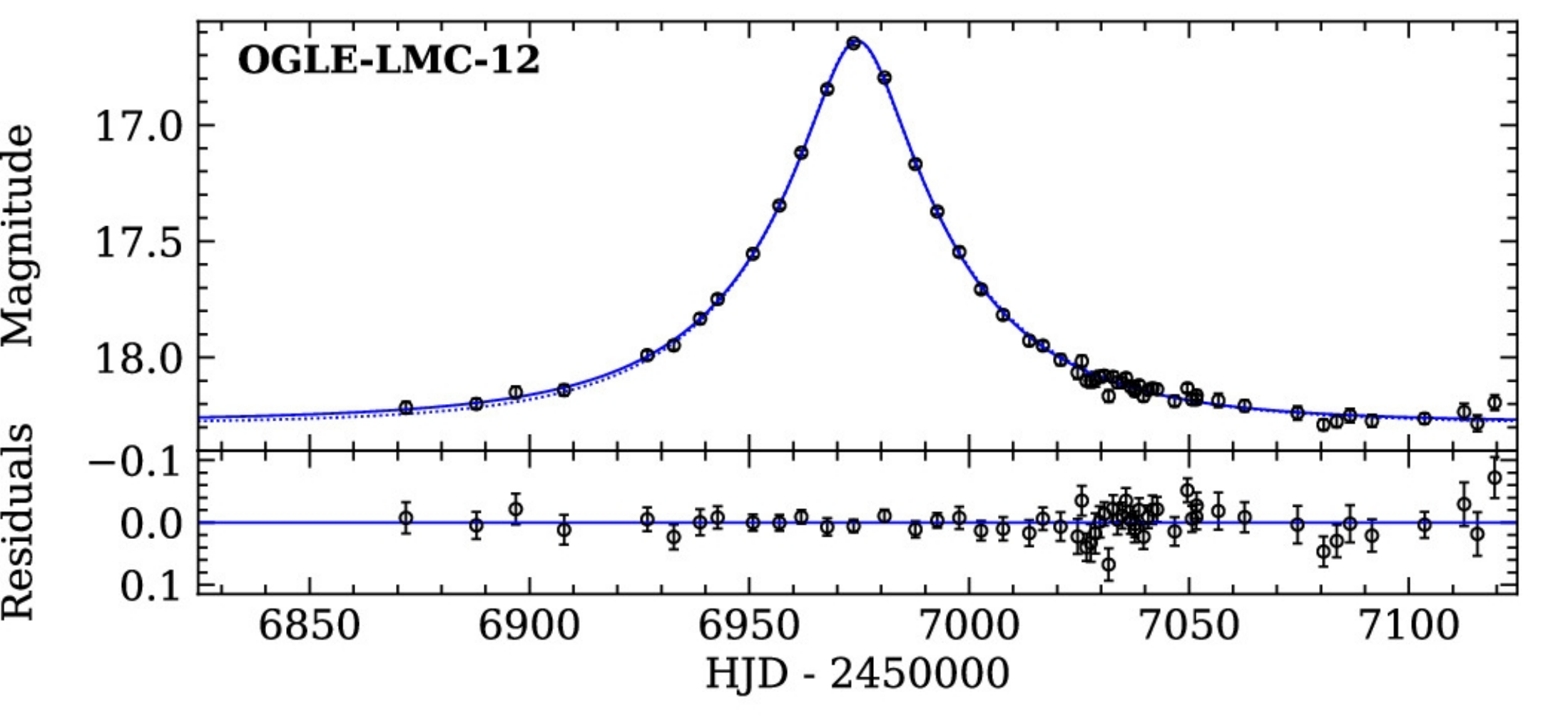}
\end{picture}
\begin{picture} (150,0) (45,-10)
\includegraphics[width=0.49\textwidth]{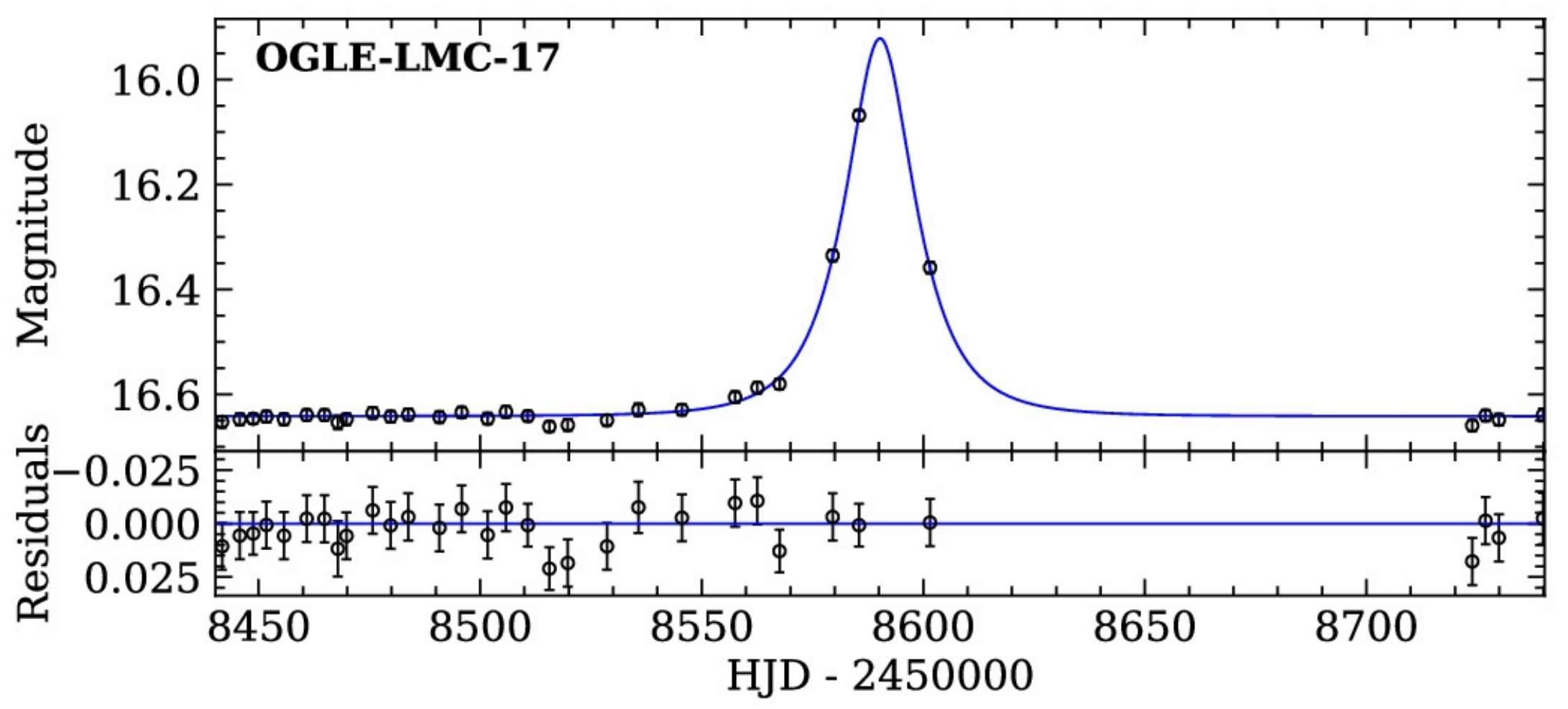}
\end{picture}
\caption{Light curves in the $I$-band adapted from Figure 4 of
 Mr\'{o}z et al. (2024b) for two candidate microlensing events.}
\label{fig4}
\end{figure}

An important factor in successfully detecting a microlensing event is the
cadence or frequency of the observations, in other words how well sampled
are the light curves.  \cite{m24a} quote the average cadence for an OGLE
field to be in the range 3-10 days.  The corresponding figure for the
MACHO fields can be calculated from data given by \cite{a00} to be in the
range 1.5-3.5 days.  This is a big difference, and comparison of the light
curves in Fig.~\ref{fig3} adapted from \cite{a00} with those in
Fig.~\ref{fig4} adapted from \cite{m24a} confirms the superiority of the
MACHO data for the definition of light curve features.  For example,
several of the OGLE light curves are only sparsely covered, and in some
cases only half complete.  Moreover, some microlensing events could
easily have fallen in the gaps between measurements.

\subsection{Microlensing event selection}

An important part of detecting microlensing events is the selection
criteria or "cuts" for deciding which light curves contain microlensing
events.  If the cuts are made too severe some events will be missed,
but if they are made too inclusive, the sample of candidate events will be
contaminated by light curves with spurious features not associated with
microlensing.  The MACHO project presented two sets of selection
criteria.  The first selection A was designed to accept only high quality
candidates, largely involving extent of observational coverage, amplitude
and cross-correlation between red and blue light curves.  Selection B was
designed to accept any light curve with a significant peak and flat
baseline, and set less stringent limits than selection A.  This resulted
in 13 and 17 events from selection A and B respectively.

The OGLE project took a very different approach to selecting microlensing
candidates.  This appears to be mainly the result of the limitions of
their observational data.  According to Table 4 of \cite{m24a} there were
only weak constraints on the number of points in the "bump", presumably
because of the sparseness of the data points in their light curves.  There
were however strong constraints on the goodness of fit to their
microlensing model, which may have been necessitated by the absence of a
second colour in the light curves.  In the crowded star fields of the LMC,
a good fit must have been hard to achieve, even for a genuine microlensing
event.  The resulting decrease in the likelihood of event detections
should be allowed for by a corresponding decrease in the estimated
detection efficiency, as discussed below.  However, failure to do this
correctly will result in an underestimate of the expected number of
microlensing events, and hence of the optical depth to microlensing.  It
would also provide an explanation as to why the OGLE group detect fewer
microlensing events than the MACHO group. 

\subsection{Detection efficiency}

It is clear from the discussion on microlensing event selection that a
significant number of microlensing events are likely to go undetected.  To
allow for this the MACHO project developed a procedure to estimate the
detection efficiency, which can be seen as the number of microlensing
events actually detected as a fraction of the number of events
consistent with the cuts or selection criteria which could have been
detected.  The possible reasons for non-detection are many and varied.
With good book keeping, such issues as times of observation and what
sources are observed can be relatively easily resolved, but even here
there can be problems such as how to deal with incomplete light curves.
More problematic issues include such effects as microlensing profile
distortion and colour changes produced by neighbouring stars in crowded
fields.  Similarly, changes in seeing can result in changes in image
overlap and hence microlensing profile from night to night.  These
problems are well illustrated in Figure 4 of \cite{a01} as mentioned
above.

The problems in calculating the detection efficiency were widely
acknowledged in the MACHO papers \citep{a00,a01a}, and are discussed in
detail by \cite{h15}.  The MACHO collaboration concluded that the only way
to proceed was by constructing a Monte Carlo type process where artificial
microlensing events were injected into typical light curves. Event
detection algorithms, including relevant cuts, were then run to measure
the percentage of successfully identified microlensing events, defined as
the photometric efficiency.  Although this procedure determined whether an
artificial microlensing event would be detected, it did not provide useful
measurements of important parameters such as the amplitude and timescale
of the event.  In addition, in order to measure the optical depth to
microlensing, an accurate knowledge of the number of source stars is
required, which can be heavily disguised in a crowded star field.  These
issues were addressed with additional extensive analysis by \cite{a01a}.

In order to calculate the detection efficiency, the photometric efficiency
must be combined with the actual survey parameters, including number of
stars observed, timescale of survey and relevant selection criteria.  The
results are usually presented as a function of detection efficiency versus
microlensing event duration.  It should be emphasised that the detection
efficiency is very sensitive to the selection criteria, and basically has
to allow for all the actual events in the survey field which are excluded
by the selection criteria.

A major problem with current approaches to detection efficiency is to find
a way to objectively verify the results of the Monte Carlo process.  There
are however in favourable circumstances ways in which this can be done.
Where there is an overlap between two surveys when the same field is
observed by both surveys at the same time, the number of common event
detections can be used as an estimate of the detection efficiency.  The
two microlensing events detected in the OGLE-II phase of the survey
\citep{w09} were both detected by the MACHO survey, but did not form part
of their sample of candidate microlensing events.  OGLE-LMC-01 was
detected as a microlensing candidate, but it was not included in the MACHO
final sample \citep{a00} as it occurred shortly after their sample time
limit.  OGLE-LMC-02 was also observed by the MACHO collaboration, but not
classified as a microlensing event, probably due to poor quality data.
There is no published record of any detections by the OGLE group of MACHO
microlensing candidates.  If the detection efficiencies of both projects
are estimated correctly, then differences in observing procedures and
detection strategies will all be taken into account, and the resulting
measurements of optical depth to microlensing for both surveys will be the
same.  As it stands, the MACHO detections reported by \cite{a00} have
largely stood the test of time in spite of intense scrutiny, and the
puzzle remains as to why the detection efficiency estimated by the OGLE
group has not been able to compensate for their lack of detections.
Another way of verifying the correctness of detection efficiency estimates
is to search for microlensing events involving known populations of stars.
In this case, the applcation of detection efficiency to the number of
microlensing candidates should give the predicted number of events.  This
test is applied to the OGLE detections in the self lensing subsection
below.

The OGLE group's approach to estimating the detection efficiency
\citep{m24b} is summed up in 3 short paragraphs, and appears to be based
on early work by \cite{w09}.  This in turn is based on the Monte Carlo
approach taken by the MACHO group \citep{a01a}, although few details are
given.  However, the entire paper devoted to the subject by the MACHO
collaboration \citep{a01a} illustrates the very complex nature of the
estimation of detection efficiency, and it is far from clear from
published work that the OGLE group has developed an adequate procedure
to tackle the problem.

Finally on the subject of detection efficiency, there are a number of
puzzling inconsistencies in the published detection efficiency diagrams.
For example, despite the fact that OGLE microlensing candidates are only
selected in the $I$-band with no colour information, and from what are
arguably inferior and less well-sampled light curves, the adopted
detection efficiency as given by \cite{m24b} appears to be about the same
as that for the MACHO candidates in \cite{a00}.  In addition, the severity
of the cuts in Table 4 of \cite{m24b} compared with those in Table 3 of
\cite{a00} does not seem to be reflected in the respective detection
efficiences, and it seems very likely that the values adopted by \cite{m24b}
are far too large.

\subsection{Self lensing}
\label{sl}

It is a welcome development that a useful sample of microlensing events
has finally been detected by the OGLE collaboration.  The breakthrough
seems to have been achieved by extending the length of the survey from
the combined OGLE-II and OGLE-III phases lasting a total of approximately
13 years from 1996 until 2009.  This combined phase resulted in the
detection of a total of 4 microlensing events, and although the addition
of OGLE-III data resulted in the survey area being increased by a
factor of 10, all detections were made in the original area covered by
OGLE-II in the central part of the LMC.  This is not surprising, as it
included the highest density of sources.  The final phase of the OGLE
programme was to combine a new run of observations designated OGLE-IV with
the existing OGLE-III data to give a continuous set of light curves
covering 10 years from 2010 to 2020.  The area covered by the new survey
was also increased from 40 deg$^2$ to 300 deg$^2$.  This survey resulted
in the detection of 13 microlensing events, all close to the original
OGLE-II fields.  It is perhaps strange that the earlier phase of the OGLE
programme, although lasting for 13 years only detected 4 microlensing
events, whereas the final phase lasting 10 years detected 13 events in
roughly the same area of sky.  Although there were presumably some small
improvements to the reduction procedures, \cite{m24b} point out that the
observing setup and reduction pipeline were similar to OGLE-III.

Perhaps the most significant conclusion of the recent results from the
OGLE survey \citep{m24a} is that the 13 candidate microlensing events can
be accounted for by self lensing.  Since the results of the MACHO survey
were published \citep{a00}, there have been a number of estimates of self
lensing towards the LMC, including stellar lenses in the Galactic disc and
in the LMC itself.  The value of optical depth to microlensing $\tau$
towards the LMC for self lensing is basically a function of Galactic
structure, and there is a reasonable consistency between published values
for $\tau$ which are mostly in the range
$\tau = 0.4 \pm 0.1 \times 10^{-7}$ \citep{a00,m04,c09}.  Comparing this
with the measured value of $\tau = 0.121 \pm 0.037 \times 10^{-7}$ from
the 13 microlensing candidates certainly suggests that if the analysis is
correct it is small enough to be consistent with self lensing.  More
worrying is why the expected number of around 40 self lensing events were
not detected.  In fact, the actual value of $\tau$ derived by \cite{m24b}
has not changed much from the results of OGLE-III which gave
$\tau = 0.16 \pm 0.12 \times 10^{-7}$.  The main difference is the large
reduction in the size of the error bars resulting from the increased
number of event detections.  The failure to detect the known background of
self lensing events at such a high signficance level suggests that there
are unidentified issues with the selection criteria, and almost certainly
with the estimation of photometric and detection efficiency.  It would
thus seem premature at this point to attempt to put any microlensing
constraints on the population of compact bodies in the Galaxy.

The question of self lensing by Milky Way disc stars can in principle be
settled by looking for anomalies in the colours or spectra of the source
stars.  The most likely position for the lens in a microlensing event is
half way between the observer and the source.  For the LMC this is
about 25kpc from Earth.  This means that red giant lenses superimposed
on the source during a microlensing event should be easy to identify,
and main sequence stars down to spectral type K5 will be of comparable
brightness to most source stars.  In fact \cite{m24b} have already
confirmed that source star colours during a microlensing event are
generally similar to the base line colour "confirming that the majority of
the observed light originates from the source star".  This means that the
lens must either be dark or much fainter than the source star.  For
disc stars acting as lenses, it should be possible to perform
spectroscopic or at least photometric follow-ups of the lensed star to see
if the spectrum or the photometry varied over the course of the event.
This is especially true for long duration events which necessarily
indicate a massive lens that in the case of a disc star should be bright
enough to be detectable.  Although it may be possible to make the case
that some models of the Milky Way and LMC appear to be consistent with the
identification of observed microlensing events with self lensing, until
there is positive identification of any stellar lenses and the detection
efficiency is reliably known this cannot be taken to rule out or even
constrain any specific population of compact bodies as components of the
Galactic halo.

A final note of caution is worth making with regards to the identification
of microlensing events with self lensing. Although it is true that
plausible models of the Milky Way and LMC appear to be consistent with the
observed microlensing events being accounted for by self lensing, the
models themselves are very uncertain, and in fact the two models adopted
for the analysis of the OGLE observations \citep{h03,c20} differ by a
factor of two in the expected number of self lensing microlensing events.
Although it may be true that given the relatively small number of
OGLE microlensing event detections there is a good case that some or even
all of them may be attributed to self lensing, there is no way that this
can be seen as any sort of limit to a proposed population of primordial
black holes in the Milky Way halo, as implied by the bullish title of the
recent {\it Nature} paper `No massive black holes in the Milky Way halo'
\citep{m24a}.  The question of whether microlensing detections are
sufficient to imply a significant population of primordial black holes or
other compact body dark matter candidates in the Galactic halo is a
separate issue, and requires a dynamic measurement of the mass profile of
the Galaxy, which we address in the following subsection.

\subsection{Halo models}

When Bohdan Paczy\'{n}ski proposed a search for microlensing events in the
light curves of Magellanic cloud stars, his clearly stated aim was to look
for compact bodies which might make up the dark matter in the Galactic
halo.  So far, the focus of this paper has been on the complex process of
detecting a well-defined sample of microlensing events to provide an
accurate measurement of the optical depth to microlensing $\tau$ towards
the LMC.  However, the final and perhaps most important question is
whether the population of compact bodies acting as lenses is consistent
with making up the dark matter.  To estimate this it is necessary to
derive a mass model of the Galaxy, with particular reference to the halo,
which typically involves the measurement of the Galactic rotation curve.

The first successful measurement of $\tau$ \citep{a00} was based on 13 to
17 microlensing events (depending on the selection criteria), and gave a
value of $\tau = 1.2 \times 10^{-7}$.  The next and most controversial
step was to convert the implied mass into a fraction of the dark matter
halo mass.  To do this, it was necessary to estimate the Milky Way
rotation curve, which was very poorly known at that time \citep{h15}.
In the event the MACHO collaboration opted to use the relatively easily
measured rotation curves of nearby spiral galaxies as plausible models for
the Milky Way.  This resulted in the adoption of a flat rotation curve,
implying a heavy halo, although they did consider other models for
comparison.  The assumption of a flat rotation curve (Model S from
\cite{a96}) implied a most likely MACHO halo fraction of 20\% with a
maximum value of 50\% at the 95\% confidence level.  Although at the time
these findings were arguably considered a disappointment by the MACHO
collaboration who were hoping to detect a 100\% MACHO halo, it was
nonetheless a remarkable result.  The discovery of a large new population
of stellar mass compact bodies which were inconsistent with any known
population of stars certainly suggested that they might make up at least a
substantial fraction of the dark matter, and it was even suggested at that
time that they might be primordial black holes \citep{a97}.

In the ensuing years there were steady improvements in the direct
measurement of the Galactic rotation curve \citep{x08,s13,b14}, and it
became clear that the Galactic rotation curve was not flat, but steadily
declined at least as far as the distance to the Magellanic Clouds.  This
had important implications for the results of the MACHO microlensing
survey, and \cite{h15} showed that it was easy to find models for the
Galactic halo which were consistent with the new rotation curve
measurements, and also with the value of optical depth to microlensing
$\tau$ derived by \cite{a00} for a 100\% MACHO halo.

In this context it is surprising to find that the OGLE results are still
interpreted with a model that is not consistent with the Milky Way
rotation curve \citep{m24a}.  Their analysis is based on the rotation
curve from \cite{c20} which is itself largely based on \cite{e19} and
which only extends to 25 kpc.  The resulting rotation curve derived by
\cite{c20} has little useful data beyond 25 kpc.  At 50 kpc, the distance
of the LMC, the circular velocity from \cite{c20}, and apparently adopted
by \cite{m24a}, is 190 km sec$^{-1}$.  This is virtually the same as for
the the heavy-halo Model S adopted by \cite{a00}, and should be compared
with the modern value of 130 km sec$^{-1}$ based on {\it Gaia} DR3 data
\citep{o24,g24}.  It is of course true to say that as the OGLE survey
resulted in so few event detections it is somewhat irrelevant as to
whether they used a correct model of the Galactic halo to estimate the
halo fraction in compact bodies.  However, if their detection algorithms
improve it is important that they should not repeat the mistaken choice of
halo model which has caused so much confusion in interpreting the MACHO
results.

\section{Discussion}
\label{dis}

This paper was prompted by a recent article in {\it Nature} \citep{m24a},
and an acompanying paper in Astrophysical Journal Supplement \citep{m24b}
which stated in strongly worded terms that there are no massive black
holes in the Milky Way halo.  The evidence to support this claim by the
OGLE collaboration was at best flimsy, and under normal circumstances
could have been left to the judgement of others.  However, it soon became
clear to us that in the absence of any critique of the arguments
presented, the conclusions of the paper were widely accepted.  It is not
a pleasant task to expose the weakness of a publication from a well
respected group such as the OGLE collaboration, but the subject is of
great topical importance and we feel it is important that both sides of
the argument are heard.

The broad structure of the OGLE argument seems to be as follows.  The
OGLE project which was set up on largely similar lines to the MACHO
project was unable to reproduce the MACHO detections.  On this basis they
concluded that the population of lenses which the MACHO group observed
did not exist.  This attempt to put theoretical constraints on a
population which has actually been detected is clearly logical fallacious.
It would seem that the only way to proceed to discredit the MACHO result
is to demonstrate that most of their detections are not microlensing
events.  This approach has been tried by a number of groups including
OGLE, but so far the sample of microlensing candidates has largely
withstood the most rigorous examination.

An alternative approach which has also been pursued by the OGLE
collaboration is to accept the reality of the microlensing events, but to
suggest that the observed events can all be explained by stellar lenses in
the Galactic disc, or self lensing in the LMC.  This does not of course
put any formal constraint on the lens population of the Galactic halo, but
does however provide an unexpected test of the reliability of the OGLE
observations.

The most recent phase of the OGLE project reports the detection of 13
microlensing events which are attributed to self lensing.  This is of
course a possible explanation, but what is not made clear in the paper is
that the expected number of disc and self lensing events is around 40.
The detection of these `known' events should be a perfect test for the
effectiveness of the OGLE reduction pipeline, with particular focus on
their detection efficiency parameter.  It is now clear that this test is
failed at a very significant level, and that as it stands their results
cannot be relied upon to give any useful constraint on the population of
primordial black holes in the Galactic halo.

Moreover, if one takes at face value the 13+3 microlensing
events\footnote{Note that, apart from the 13 events obtained by the online
search pipeline, there were 3 more events found offline `by hand', which
should also be taken into account in this discussion.} from OGLE as
arising from primordial black holes with a Thermal History Model mass
function \citep{Carr21,Carr24}, and computes the expected average number
of events given the observed number according to Poisson statistics
\citep{g24}, one can accommodate 80\% (100\%) of all the DM in the halo in
the form of PBH, at 95\% (99\%) confidence level.\footnote{The calculation
of $N_{\rm obs}^*$ from Fig.3a of~\cite{g24} gives a factor 7 (10)
reduction in the OGLE constraints at 95\% (99\%) c.l.} Note that these
results arise without even changing either the efficiency function or the
rotation curves assumed by the OGLE collaboration. 

\section{Conclusions}

We summarize here our findings:

\begin{enumerate}

\item There is strong evidence that a population of stellar mass
primordial black holes have been detected in the halos of gravitationally
lensed galaxies and clusters, consistent with the detection of compact
bodies by the MACHO group in the Milky Way halo.

\item A large part of this paper has been devoted to demonstrating the
insecurity of OGLE claims to have ruled out a significant population of
compact bodies in the Milky Way halo.  This is based on a detailed
examination of the reliability of their microlensing event detection
procedures, which we show to be inadequate or incorrect in many respects.

\item Failure by the OGLE project to detect the population of compact
bodies observed by the MACHO project does not put any constraint on a
population of primordial black holes in the Galaxy halo.  The 13 events
which they did detect should be compared with the 40 events expected from
known stellar populations, and raises questions about the reliability of
their estimates of detection efficiency.

\item  The 13 microlensing events detected by the OGLE project are
consistent with the expected number of events for a PBH halo for a Thermal
History mass function.

\item The only way to challenge the MACHO results is to show that their
candidates are not microlensing events.  Despite intensive efforts this
has so far proved unsuccessful.

\item The statement by the OGLE collaboration in the title of their Nature
paper {\it No massive black holes in the Milky Way halo} is without
foundation.

\end{enumerate}

\section*{Acknowledgements}

J.G.B. acknowledges support from the Spanish Research Project
PID2021-123012NB-C43 [MICINN-FEDER], and the Centro de Excelencia Severo
Ochoa Program CEX2020-001007-S at IFT.

\section*{Data Availability}

The data upon which this paper is based are all publicly available and are
referenced in the text, with footnotes to indicate online archives where
appropriate.


\bsp	
\label{lastpage}

\begin{thebibliography}{}

\bibitem[\protect\citeauthoryear{Alcock et al.}{1993}]{a93}
 Alcock C. et al., 1993, Nature, 365, 621
\bibitem[\protect\citeauthoryear{Alcock et al.}{1996}]{a96}
 Alcock C. et al., 1996, ApJ, 461, 84
\bibitem[\protect\citeauthoryear{Alcock et al.}{1997}]{a97}
 Alcock C. et al., 1997, ApJ, 486, 697
\bibitem[\protect\citeauthoryear{Alcock et al.}{1999}]{a99}
 Alcock C. et al., 1999, PASP, 111, 1539
\bibitem[\protect\citeauthoryear{Alcock et al.}{2000}]{a00}
 Alcock C. et al., 2000, ApJ, 542, 281
\bibitem[\protect\citeauthoryear{Alcock et al.}{2001}]{a01}
 Alcock C. et al., 2001, ApJ, 552, 582
\bibitem[\protect\citeauthoryear{Alcock et al.}{2001a}]{a01a}
 Alcock C. et al., 2001a, ApJ, 136, 439
\bibitem[\protect\citeauthoryear{Bhattacharjee et al.}{2014}] {b14}
 Bhattacharjee P., Chaudhury S., Kundu S., 2014, ApJ, 785, 63
\bibitem[\protect\citeauthoryear{Calcino et al.}{2018}]{c18}
 Calcino J., Garc\'{i}a-Bellido J., Davis T.M., 2018, MNRAS, 479, 2889
\bibitem[\protect\citeauthoryear{Cautun et al.}{2020}]{c20}
 Cautun M. et al., 2020, MNRAS, 494, 4291
\bibitem[\protect\citeauthoryear{Calchi Novati et al.}{2009}]{c09}
 Calchi Novati S. et al., 2009, MNRAS, 400, 1625
\bibitem[\protect\citeauthoryear{Carr et al.}{2021}]{Carr21}
 Carr B. et al., 2021, Phys.Dark Univ. 31, 100755
\bibitem[\protect\citeauthoryear{Carr et al.}{2024}]{Carr24}
 Carr B. et al., 2024, Phys.Rept. 1054, 1
\bibitem[\protect\citeauthoryear{Eilers et al.}{2019}]{e19}
 Eilers A.-C., Hogg D.W., Rix H.-W., Ness M.K., 2003, ApJ, 871, 120
\bibitem[\protect\citeauthoryear{Garc\'ia-Bellido \& Hawkins}{2024}]{g24}
 Garc\'ia-Bellido J., Hawkins M.R.S., 2024, Universe, 10, 449
\bibitem[\protect\citeauthoryear{Han \& Gould}{2000}]{h03}
 Han C., Gould A., 2003, ApJ, 592, 172
\bibitem[\protect\citeauthoryear{Hawkins}{2015}]{h15}
 Hawkins M.R.S., 2015, A\&A, 575, A107
\bibitem[\protect\citeauthoryear{Mancini et al.}{2004}]{m04}
 Mancini L. et al., 2004, A\&A, 427, 61
\bibitem[\protect\citeauthoryear{Mr\'{o}z et al.}{2024a}]{m24a}
 Mr\'{o}z P. et al., 2024a, Nature, 632, 749
\bibitem[\protect\citeauthoryear{Mr\'{o}z et al.}{2024b}]{m24b}
 Mr\'{o}z P. et al., 2024b, ApJS, 273, 4
\bibitem[\protect\citeauthoryear{Mr\'{o}z et al.}{2024b}]{m19}
 Mr\'{o}z P. et al., 2019, ApJS, 244, 29
\bibitem[\protect\citeauthoryear{Ou et al.}{2024}]{o24}
 Ou X., Eilers A.-C., Necib L., Frebel A., 2024, MNRAS, 528, 693
\bibitem[\protect\citeauthoryear{Paczy\'{n}ski}{1986}]{p86}
 Paczy\'{n}ski B., 1986, ApJ, 304, 1
\bibitem[\protect\citeauthoryear{Sofue}{2013}]{s13}
 Sofue Y., 2013, PASJ, 65, 118
\bibitem[\protect\citeauthoryear{Witt \& Mao}{1994}]{w94}
 Witt H.J., Mao S., 1994, ApJ, 430, 505
\bibitem[\protect\citeauthoryear{Witt \& Atrio-Barandela}{2019}]{w19}
 Witt H.J., Atrio-Barandela , 2019, ApJ, 880, 152
\bibitem[\protect\citeauthoryear{Wyrzykowski et al.}{2009}]{w09}
 Wyrzykowski \L. et al., 2009, MNRAS, 397, 1228
\bibitem[\protect\citeauthoryear{Wyrzykowski et al.}{2011}]{w11}
 Wyrzykowski \L. et al., 2011, MNRAS, 413, 493
\bibitem[\protect\citeauthoryear{Xue et al.}{2008}]{x08}
 Xue X.X. et al., 2008, ApJ, 684, 1143

\end{thebibliography}
\end{document}